\documentclass[12pt]{article}

\catcode`\@=11

\@addtoreset{equation}{section}
\renewcommand{\theequation}{\thesection.\arabic{equation}}

\newcommand{\address}[1]{\vbox{\let\\=\cr \normalsize \vskip 1em
  \lineskip\normallineskip \halign{\hfil##\hfil\crcr#1\crcr}}}

\newcommand{\topic}[1]{\medskip\noindent\textit{#1}\medskip\noindent}

\newcommand{\mbb}[1]{\mathrm{#1}}
\newcommand{\mfs}[1]{\mathcal{#1}}

\DeclareFontFamily{U}{cmss}{}
\DeclareFontShape{U}{cmss}{bx}{n}{<-> cmssbx10}{}
\DeclareMathAlphabet{\mathsb}{U}{cmss}{bx}{n}

\newcommand{\newaside}[2]{%
  \newenvironment{#1}[1][\relax]{\@beginaside{#2}{##1}}{\@endaside}}

\newcommand{\@beginaside}[2]{%
  \footnotesize\normalfont\trivlist 
  \item[\hskip\labelsep{\itshape #1\/}:]%
  \ifx\relax#2  \else  ({\bfseries #2\/})  \fi}
\newcommand{\@endaside}{\endtrivlist}

\newaside{example}{Example}
\newaside{exercise}{Exercise}
\newaside{solution}{Solution}
\newaside{remark}{Remark}

\newenvironment{eqtableau}[1]{\@begineqtableau{#1}}{\@endeqtableau}

\def\@begineqtableau#1{
  \vcenter\bgroup\openup1\jot
    \mathsurround=0pt  \everymath={\displaystyle}
    \count0=#1  \toks0={\strut}  \toks1={##}  \let\\=\crcr
    \ifnum\count0 > 0
      \loop  \advance\count0 by -1
        \@ApndToks{\toks0}{$\hfil\the\toks1$&${}\the\toks1\hfil$}
      \ifnum\count0 > 0
        \@ApndToks{\toks0}{&}
      \repeat
    \else  \@ApndToks{\toks0}{\hfil$\the\toks1$\hfil}  \fi
    \edef\@act{\noexpand\ialign\bgroup\the\toks0\noexpand\crcr}  \@act}
\def\@endeqtableau{\crcr\egroup\egroup}

\def\@ApndToks#1#2{\edef\@act{\noexpand#1={\the#1 #2}}  \@act}

\newenvironment{eqset}{\@begineqset}{\@endeqset}
\newcommand{\interject}[1]{%
  \noalign{\vskip2\jot\vbox{\noindent #1}\vskip2\jot}}

\def\@begineqset{\begingroup  \mathsurround=0pt  \let\\=\@eqsetcr%
  \stepcounter{equation}\def\@currentlabel{\p@equation\theequation}%
  $$\everycr={}  \everymath={\displaystyle}  \openup\jot
    \tabskip=\@centering  \halign to\displaywidth\bgroup
    \hfil$##$\tabskip=0pt&${}##$\hfil\tabskip=\@centering&\relax
      \llap{##}\tabskip=0pt\crcr}
\def\@endeqset{\egroup\global\advance\c@equation by -1$$
  \endgroup\@ignoretrue}

\def\@eqsetcr{&\if@eqnsw\@eqnnum\stepcounter{equation}\fi
  \global\@eqnswtrue\crcr}

\newcommand{\dens}[2][1]{{
  \mathsurround=0pt  \everymath={\displaystyle}  \toks0={}
  \setbox0=\hbox{\lower 4.30554pt\hbox{$\mathchar"0365$}}  \dp0=0pt
  \count0=#1  \ifnum\count0 < 0  \multiply\count0 by -1  \fi
  \loop  \advance\count0 by -1  
    \@ApndToks{\toks0}{\copy0\crcr} 
  \ifnum\count0 > 0
    \@ApndToks{\toks0}{\noalign{\kern -1pt\nointerlineskip}}
  \repeat
  \count0=#1  \skip3=0pt plus 3fil  \skip1=0pt plus 1fil
  \ifnum\count0 > 0
    \vbox{\ialign{\hskip\skip3##\hskip\skip1\crcr
      \the\toks0\noalign{\nointerlineskip}${#2}$\crcr}}
  \else
    \vtop{\ialign{\hskip\skip1##\hskip\skip3\crcr
      ${#2}$\crcr\noalign{\nointerlineskip\kern 2pt}\the\toks0}}
  \fi}\vphantom{#2}}

\newcommand{\tensor}[3]{\relax}

\def\tensor#1<#2#3#4>{{\let\\=\relax  \@beginTensorScripts
  \mathsurround=0pt  \everymath={\displaystyle}
  \futurelet\@@next\@tensorStem #2\\
  \hbox{$\,{\copy1#1} \!#2#3\bgroup\@tensorLoop #4\\ \,$}}
  \@endTensorScripts  \catcode`\ =9  \catcode`\^^M=9
  \futurelet\CGPG@@next\@nextTensor}

\def\@nextTensor{\catcode`\ =10  \catcode`\^^M=5
  \let\CGPG@next=\relax
  \ifx<\CGPG@@next  \let\CGPG@next=\tensor  \fi
  \ifx^\CGPG@@next  \let\CGPG@next=\tensor  \fi
  \ifx_\CGPG@@next  \let\CGPG@next=\tensor  \fi  \CGPG@next}

\newcommand{\tensorscripts}[1]{\@beginTensorScripts{#1}\@endTensorScripts}
\newcommand{\tensorbox}[1]{\@beginTensorScripts\hbox{$\displaystyle 
  #1$}\@endTensorScripts}

\newdimen\CGPG@subscriptDepth
\newcommand{\@beginTensorScripts}{{\usefont{OMS}{cmsy}{m}{n}%
  \global\dimen\CGPG@subscriptDepth=\fontdimen16\font 
  \fontdimen16\font=\fontdimen17\font}}
\newcommand{\@endTensorScripts}{{\usefont{OMS}{cmsy}{m}{n}%
  \fontdimen16\font=\dimen\CGPG@subscriptDepth}}

\def\@tensorStem#1\\{
  \setbox0=\hbox{$#1$}
  \ifx\dens\@@next  \setbox1=\hbox{$\vphantom{\@lastToken #1\\}$}
  \else             \setbox1=\hbox{$\vphantom{#1}$}  \fi}

\def\@lastToken#1\\{\@@lastToken#1\relax\\} 
\def\@@lastToken#1#2\\{
  \ifx\relax#2  #1  \else  \@@lastToken#2\\  \fi}

\def\@tensorLoop{\futurelet\@@next\@@tensor}
\def\@@tensor{
  \ifx^\@@next   \let\@next=\@tensorSup  \else
  \ifx_\@@next   \let\@next=\@tensorSub  \else
  \ifx:\@@next   \let\@next=\@tensorCon  \else
  \ifx\\\@@next  \let\@next=\@tensorEnd  \else
                 \let\@next=\@tensorChr  \fi\fi\fi\fi  \@next}
\def\@tensorSup#1{\egroup\copy1^\bgroup\@tensorLoop}
\def\@tensorSub#1{\egroup\copy1_\bgroup\@tensorLoop}
\def\@tensorChr#1{{#1}\@tensorLoop}
\def\@tensorEnd#1{\egroup}
\def\@tensorCon#1#2{
  \ifx_#2  \let\@next=\@tensorCSb  \else
  \ifx^#2  \let\@next=\@tensorCSp  \fi\fi  \@next}
\def\@tensorCSb{\egroup_\bgroup\@tensorLoop}
\def\@tensorCSp{\egroup^\bgroup\@tensorLoop}

\newcommand{\overcirc}[1]{\vbox{\mathsurround=0pt  
  \skip1=0pt plus 1 fil  \skip3=0pt plus 3fil
  \ialign{##\crcr$\hskip\skip3\scriptstyle\circ\hskip\skip1$\crcr
    \noalign{\kern1pt\nointerlineskip}$\displaystyle{#1}$\crcr}}
  \vphantom{#1}}

\newcommand{\grad}[0]{\nabla}
\newcommand{\weq}[0]{\approx}

\newcommand{\implies}[0]{\Rightarrow}

\newcommand{\order}[0]{\mfs{O}}

\newcommand{\tsfrac}[2]{{\textstyle\frac{#1}{#2}}}

\newcommand{\abs}[1]{\left| #1 \right|} 

\newcommand{\pbrack}[2]{\left\{ #1, #2 \right\}}

\newcommand{\bra}[2][<]{\CGPG@enclosed #1|{#2}}
\newcommand{\ket}[2][>]{\CGPG@enclosed |#1{#2}}
\newcommand{\braket}[3][<>]{\CGPG@enclosed 
  #1{\left.#2\vphantom{#3}\right|#3}}
\newcommand{\bopk}[4][<>]{\CGPG@enclosed 
  #1{#2 \CGPG@enclosed ||{\vphantom{#2}#3\vphantom{#4}} #4}}
\newcommand{\expect}[2][<>]{\CGPG@enclosed #1{#2}}

\newcommand{\norm}[2][\|]{\CGPG@enclosed #1#1{#2}}
\newcommand{\iprod}[3][<>]{\CGPG@enclosed #1{#2, #3}}

\newcommand{\CGPG@enclosed}[3]{\left#1 #3 \right#2}

\renewcommand{\Re}[0]{\mbb{R}}

\catcode`\@=12

\newcommand{\GamBar}[0]{\overline\Gamma}
\newcommand{\OmBar}[0]{\overline\Omega}
\newcommand{\SBar}[0]{\overline S}

\begin{document}

\title  {Midi-Superspace Quantization of Non-Compact Toroidally 
         Symmetric Gravity}
\author {Christopher Beetle\\
  \address{Center for Gravitational Physics and Geometry\\
           Department of Physics, The Pennsylvania State University\\
           University Park, PA 16802}}
\date   {January 29, 1998}
\maketitle

\begin{abstract}
We consider the quantization of the midi-superspace associated with a 
class of spacetimes with toroidal isometries, but without the compact 
spatial hypersurfaces of the well-known Gowdy models.  By a symmetry 
reduction, the phase space for the system at the classical level can be 
identified with that of a free massless scalar field on a fixed background 
spacetime, thereby providing a simple route to quantization.  We are then 
able to study certain non-perturbative features of the quantum 
gravitational system.  In particular, we examine the quantum geometry of 
the asymptotic regions of the spacetimes involved and find some 
surprisingly large dispersive effects of quantum gravity.
\end{abstract}

\section{Introduction}

Any quantum theory of gravity must deal with two sets of difficulties 
inherent in classical general relativity.  The first is the 
diffeomorphism invariance of the theory, from which it follows that 
there is generally no fixed spacetime geometry on which quantization 
can be performed.  Indeed, the geometry itself is the dynamical 
quantity we study.  The second set of difficulties arise because the 
field equations of general relativity are highly nonlinear and 
difficult to solve.  These features raise both technical and 
conceptual problems with the formulation of general relativity as a 
physical theory in the conventional sense.  To understand the nature 
of these problems, and perhaps their solutions, it is useful to 
consider simpler models which also exhibit them.  One class of such 
models can be found by requiring the full theory of general relativity 
to satisfy various symmetry properties, thereby reducing the number of 
possible spacetimes while leaving at least the diffeomorphism 
invariance essentially intact.

The idea of using symmetry reduction to simplify the solution of 
Einstein's equations is nearly as old as general relativity itself.  In 
fact, all the known solutions to date have been found under the assumption 
of one kind of symmetry or another.  The application of such 
simplifications to the problem of quantum gravity, however, seems to have 
begun with the so-called mini-superspace systems.  These systems consist 
of general relativity, perhaps coupled with some matter fields, together 
with the requirement of so much symmetry that the number of independent 
degrees of freedom becomes finite.  Since mini-superspace systems derive 
from general relativity, they do not depend on any fixed background 
spacetime structure.  On the other hand, since they retain only a finite 
number of degrees of freedom, they do not capture the non-linear 
\textit{field}-theoretic complexity of a complete gravitational theory.  
To incorporate both of these features, we need to consider some less 
restrictive (i.e., less symmetric) models.

One of the simplest examples of a symmetry reduced gravitational system 
arises from the dimensional reduction of 3+1-dimensional general 
relativity with respect to a (spacelike) hypersurface orthogonal Killing 
vector field.  It is well known that such a system is mathematically 
equivalent to 2+1-dimensional general relativity coupled with a free, 
massless scalar field \cite{exact, ABS}.  Although the degrees of freedom 
of 2+1-dimensional gravity are topological in nature and therefore finite 
in number, the reduced model still possesses an infinite number of degrees 
of freedom which are characterized by the excitations of the scalar field.  
Unfortunately, despite the comparative simplicity of these dimensionally 
reduced systems, they remain somewhat difficult to study in practice.  The 
problem lies in the question of global existence of solutions to the 
2+1-dimensional field theory in the presence of gravity.  To resolve this 
issue, it is sufficient to posit the existence of a second, independent, 
hypersurface orthogonal Killing field on the original 3+1-dimensional 
spacetime.  One can then show that the resulting symmetry of the 
2+1-dimensional spacetime effectively \textit{decouples} the dynamics of 
the scalar field from that of the gravitational field in the sense that 
one can solve for the scalar field without any reference to the 
\textit{physical} spacetime geometry.  Rather, it becomes possible to 
solve for the scalar field on a fictitious, fixed background spacetime and 
then, afterwards, solve for the gravitational degrees of freedom in terms 
of a given scalar field configuration.  This procedure demonstrates 
explicitly the global existence of solutions to the reduced theory.  Thus, 
the assumption of two Killing vectors removes most of the technical 
difficulty associated with the diffeomorphism invariance of general 
relativity while leaving its field-theoretic properties intact.  One might 
hope, therefore, to gain some further insight into the nature of quantum 
gravity through the study of such systems.

The most familiar examples of 3+1-dimensional spacetimes with a pair of 
independent hypersurface orthogonal Killing fields are those of 
Einstein-Rosen waves and the Gowdy models.  The first of these describes 
asymptotically flat\footnote{Note that due to the translational symmetry 
of the Einstein-Rosen spacetimes, they cannot be asymptotically flat in 
the usual 3+1-dimensional sense.  Rather, the manifold of orbits of the 
translational Killing field is required to be flat in an appropriate 
2+1-dimensional sense \cite{ABS, varad, oneKVF}.} metrics on $\Re^4$ which 
are cylindrically symmetric.  The Gowdy models describe 3+1-dimensional 
spacetimes with compact spatial topology --- $T^3$, $S^2 \times S^1$ or 
$S^3$ --- which have a toroidal symmetry group, $U(1) \times U(1)$.  The 
quantization of these systems is not a new area of research.  The quantum 
Einstein-Rosen model, for example, was described by Kucha\v{r} 
\cite{kucCyl} as early as 1971.  More recently, both of these models have 
been analyzed \cite{pier1, pier2} from a more rigorous standpoint, with 
careful attention paid to the definition of the phase space and to certain 
subtleties of the quantum theory.  The present paper performs a similar 
analysis for a third class of spacetimes which has only recently been 
introduced by Schmidt \cite{schmidt}.  The spacetimes in question have 
topology $\Re^2 \times T^2$ and their Killing fields are assumed to have 
compact, toroidal orbits.  Although the toroidal symmetry which the 
Schmidt model shares with the Gowdy models will give rise to certain 
similarities between the two, the global structures of the spacetimes 
involved are very different.  In particular, the spacetimes considered 
here do not possess compact spatial hypersurfaces as in the Gowdy models 
\cite{pier2}.  Furthermore, they are also not asymptotically flat as in 
the Einstein-Rosen model \cite{varad}.  In the previous cases, these 
characteristics are used in part to specify the phase space of the system 
and are therefore closely connected with the final theory.  In our case, 
however, we will see that it is possible to specify the phase space 
completely without initially imposing such restrictions on the spacetime 
geometry.  Instead, the structure of the phase space will be motivated 
only by the spacetime description of the model and by certain analytical 
requirements.

Since the reduced Schmidt model can be identified with a free scalar field 
on a fixed, 2+1-dimensional background spacetime, its quantization is 
relatively straightforward.  Once that quantization is accomplished, 
however, one can express the original gravitational variables of the 
system in terms of the quantum scalar field and thereby obtain a 
non-perturbative quantum gravitational theory.  It is also possible to 
regard the reduced model of the scalar field as a quantum field theory in 
its own right by neglecting the coupling of matter to gravity.  Therefore, 
by comparing these two theories, one might reasonably hope to gain some 
insight into how the introduction of gravity will affect our current 
understanding of quantum field theory.  For example, it is often suggested 
that a quantum field theory which incorporates gravity will come equipped 
with a natural cut-off at roughly the Planck scale which will alter the 
dynamics of field excitations with trans-Planckian energies.  In this 
model, we will have a concrete example of a such a theory.  We will see 
that, although there is no natural cut-off, there are some surprising 
quantum effects in the presence of high-frequency excitations of the 
scalar field.  These effects suggest that the classical, metric structure 
of spacetime, which is a good low-energy approximation, breaks down at 
high field energies.  The dynamics of the quantum system do remain well 
defined in that regime, but the notion of a classical spacetime metric --- 
whether fixed or dynamical --- becomes a poor approximation to the full 
theory.

The outline of the paper is as follows.  In the second section, we 
describe the Hamiltonian formulation of the Schmidt model.  We pay 
special attention to the boundary conditions which we impose on the 
physical fields and to the deparameterization process that isolates 
the true degrees of freedom of the theory.  This procedure yields an 
unconstrained reduced phase space for the system which can be 
identified with that of a free scalar field on a certain background 
spacetime.  In the third section, we define a quantum analog of this 
reduced system using a standard K\"ahler quantization scheme.  We are 
then able to use the model to examine certain geometrical questions 
and develop some intuition about quantum gravity in general.  In the 
fourth and final section, we summarize the results of the previous two 
and raise some questions for future investigation.

Lastly, we should explain some of our notational conventions.  
Throughout this paper, the speed of light will be taken to be unity: 
$c = 1$.  However, since we will be interested in comparing situations 
in which different physical effects are taken into account, we will 
not do the same with the gravitational constant $G$, or with Planck's 
constant $\hbar$.

\section {Hamiltonian Formulation}

\subsection{The Midi-Superspace}

Let us begin with a precise definition of the system we will study.  The 
spacetimes of the Schmidt model \cite{schmidt} are topologically $\Re^2 
\times T^2$ and are required to support a pair of independent spacelike, 
hypersurface-orthogonal Killing vector fields.  We will assume here that 
the orbits of the Killing fields are 2-tori, so the manifold of orbits 
will be topologically $\Re^2$.  It then follows from these symmetry 
conditions that the metric $\tensor^4<g_ab>$ may be written as a sum of 
two pieces
\begin{equation}
  \tensor^4<g_ab> = \tensor^2<g_ab> + \tau\,\tensor<\sigma_ab>.
\end{equation}
Here, $\tensor<\sigma_ab>$ is a flat metric with unit total volume on a 
toroidal orbit of the symmetry group, while $\tensor^2<g_ab>$ is the 
metric on the 2-manifolds orthogonal to the orbits.  We use $\tau$ to 
denote the scale factor for the metric on the toroidal orbits in 
anticipation of its eventual role as the time parameter of the reduced 
theory.  It is possible to use $\tau$ as a time parameter since, as in the 
Gowdy model \cite{gowdy}, the symmetry properties imply that $\tau$ must 
have a timelike gradient.

The analysis presented below begins by considering the quotients 
of these spacetimes by only one of their Killing fields.  It was shown in 
\cite{ABS} that, after a conformal rescaling of its metric by $\tau$, any 
one of these quotient manifolds will define a solution of 2+1-dimensional 
general relativity coupled with a zero rest-mass scalar field.  We may 
therefore consider a quotient \textit{spacetime}, $M$, which is 
topologically $\Re^2 \times S^1$ and supports a (nowhere-vanishing) 
spacelike, hypersurface-orthogonal Killing field with closed orbits which 
we denote by $\sigma^a$.  To simplify the following discussion, we will 
work entirely within the context of this 2+1-dimensional theory.  We will 
also assume, for convenience, that both the manifold structure and all the 
fields we consider here are smooth ($C^\infty$).
 
Due to the hypersurface-orthogonality of $\sigma^a$, we can write the 
2+1-dimensional metric as
\begin{equation}\label{2+1met}
  g_{ab} = h_{ab} + \tau^2 \, \grad_a \sigma \, \grad_b \sigma.
\end{equation}
Here, $h_{ab} := \tau\tensor^2<g_ab>$ is the metric on the 2-manifolds 
orthogonal to $\sigma^a$, and $\sigma$ is the angular coordinate conjugate 
to $\sigma^a$ (i.e., $\grad_a \sigma = \tau^{-2} \, g_{ab} \sigma^b$).  
Because we assumed above that the Killing fields of the original 
3+1-dimensional spacetimes had toroidal orbits, it follows that the space 
$O$ of orbits of $\sigma^a$ will be topologically $\Re^2$ and will inherit 
a manifold structure from $M$.  Furthermore, since $\sigma^a$ is a 
spacelike Killing field, $h_{ab}$ will give rise to a metric of signature 
$(-, +)$ on $O$ which we will also denote by $h_{ab}$.  Finally, the 
function $\tau$ and the scalar field $\psi$ must be Lie-dragged by 
$\sigma^a$ and will therefore also restrict to $O$.

To begin the Hamiltonian analysis of this system, we need to introduce 
a time structure.  It is simplest to do this directly on the manifold 
$O$ of orbits of $\sigma^a$.  Thus, we choose a foliation of this 
2-manifold by spacelike lines of constant $t$, and pick a transverse 
dynamical vector field $t^a = N n^a + N^z \hat z^a$.  Here, $n^a$ is 
the unit, future-pointing, timelike normal to the foliation, and $\hat 
z^a$ is a unit-vector field on each of its leaves.  If we now pick a 
coordinate $z$ on a single leaf of the foliation, we can carry it to 
all of the others using $t^a$.  In this way, we arrive at the 
usual dynamical decomposition of the metric 
\begin{equation}\label{1+1met}
  h_{ab} = \left( -N^2 + (N^z)^2 \right) \grad_a t \,\grad_b t + 
  2N^z \,\grad_{(a}t \,\grad_{b)}z + e^\gamma \,\grad_a z \,\grad_b z.
\end{equation}
The quantities $N$, $N^z$ and $\gamma$ are all functions of $t$ and 
$z$.  Similarly, the functions $\tau$ and $\psi$ can now be expressed 
as functions of the $(t, z)$-coordinates on $O$.

The midi-superspace we have built therefore includes five real-valued 
functions on $\Re^2$: the lapse $N$, the (norm of the) shift $N^z$, 
the metric functions $\gamma$ and $\tau$, and the scalar field $\psi$.  
These functions will be required to satisfy the Einstein-Klein-Gordon 
field equations\footnote{Note that as in \cite{pier1} we have chosen 
the normalization of $\psi$ which is most consistent with the 
reduction from 3+1 to 2+1 dimensions.  This accounts for the unusual 
normalization of Einstein's equation in eq.~\ref{EKGeq}.  Strictly 
speaking, the physical scalar field is given by $\Phi := 
\psi/\sqrt{8\pi G}$, where $G$ is Newton's constant in 2+1 
dimensions.} 
\begin{equation}\label{EKGeq}
  G_{ab} = T_{ab} \quad\mbox{and}\quad g^{ab}\,\grad_a\grad_b\,\psi = 0,
\end{equation}
where $G_{ab}$ is the Einstein tensor of $g_{ab}$ and $T_{ab}$ is the 
usual stress-energy tensor of the massless scalar field $\psi$ 
\begin{equation}
  T_{ab} = \grad_a \psi \, \grad_b \psi - {\tsfrac{1}{2}} g_{ab} 
    \left( g^{cd} \, \grad_c \psi \, \grad_d \psi \right).
\end{equation}
To make sense of these as differential equations on $(N, N^z, \gamma, 
\tau, \psi)$, we need only observe that we can now construct a global 
coordinate system $(t, z, \sigma)$ on $M$.

To finish the construction of our midi-superspace, we still need to 
specify the boundary conditions for the fields it comprises.  We are, 
however, in a somewhat unusual situation in this regard.  The 
spacetimes we construct have two disjoint asymptotic regions in which 
we must specify the fall-off conditions for the fields.  In 
particular, they do not have compact spacelike hypersurfaces.  There 
is also no appropriate sense of asymptotic flatness in either of the 
asymptotic regions.  We will therefore need to use some other criteria 
to decide what the asymptotic values of the fields should be and how 
quickly they should approach them.  The criteria we will choose are 
very closely tied to the phase space formulation of the theory to be 
discussed in the next subsection.  Nevertheless, for the sake of 
completeness, we will specify here the fall-off conditions for the 
fields introduced so far.  As $z \to \pm\infty$, we require
\begin{equation}\label{mspFall}
  \begin{eqtableau}{2}
    \gamma &\to \gamma_\pm(t) + \order(z^{-1}) &\qquad
    N      &\to N_\pm(t) + \order(z^{-1})      \\
    \tau   &\to \tau_\pm(t) + \order(z^{-1})   &\qquad
    N^z    &\to \order(z^{-1})                 \\
    \psi   &\to \order(z^{-1}).                &&\\
  \end{eqtableau}
\end{equation}
The notation $f_\pm(t)$ indicates that the asymptotic values of $f$ 
can take any value, and are not a priori fixed quantities.  Also, the 
expression $\order(z^{-n})$ denotes any function $f(z, t)$ such that 
$z^n f(z, t)$, $z^{n+1} f'(z, t)$ and $z^{n+2} f''(z, t)$ all have 
finite limits as $z$ becomes infinite at fixed $t$.

The fall-off conditions in eq.~\ref{mspFall} for $\gamma$, $\tau$ and 
$N$ are about as weak as possible; they are only required to approach 
their limits in a reasonably uniform manner.  The conditions on $\psi$ 
and $N^z$ are somewhat more restrictive in that they require these 
functions to approach \textit{specific} limits, namely zero.  In the 
case of $\psi$, this fall-off condition implies that $\psi$ is square 
integrable, and therefore that its Fourier transform will exist.  This 
condition is usually imposed, even in ordinary Minkowski-space field 
theories, in order to avoid infrared divergences in the quantum field.  
This restriction is therefore justifiable on physical grounds.  The 
reason for the seemingly undesirable restriction on the asymptotic 
value of $N^z$ can best be seen within the phase space formulation to 
be discussed below.  Because of this, we will reserve its discussion 
for the next subsection.

\subsection{The Phase Space}

The usual Einstein-Hilbert action for 2+1-dimensional general 
relativity coupled to a scalar field may be written as 
\begin{equation}\label{stdAct}
  S[g, \psi] = \frac{1}{16\pi G} \int_M d^3x \,\sqrt g \, \left( R - 
  g^{ab} \,\grad_a \psi \,\grad_b \psi \right) + \frac{1}{8\pi G} 
  \oint_{\partial M} d^2x \,\sqrt h \, K.
\end{equation}
Here, $R$ represents the scalar curvature of $g_{ab}$.  The second 
integral is taken over the asymptotic boundary $\partial M$ of the 
space time, and $h$ and $K$ are respectively the determinant of the 
induced metric and the trace of the extrinsic curvature on that 
surface.  However, for the class of metrics we have chosen, it turns 
out that $K$ vanishes identically on $\partial M$.  We may therefore 
drop it from the action.  From the phase space point of view, although 
the second term is usually needed to ensure the functional 
differentiability of the action, in this case the action is already 
differentiable in its absence.  We will be able to see this below.

To pass to the Hamiltonian formulation of this theory, we make use of 
the dynamical decomposition of the metric described in 
eq.~\ref{1+1met}.  Then, by direct computation, we can find the scalar 
curvature of the metric \ref{2+1met} and use it to perform the 
Legendre transformation to momentum phase space.  As a result, the 
action takes the standard form
\begin{equation}\label{psAct}
  S = \frac{1}{16\pi G} \int dt\, \left( \int dz\, \left[ 
    p_\gamma \dot\gamma + p_\tau \dot\tau + p_\psi \dot\psi 
    \right] - C[N] - C_z[N^z] \right), 
\end{equation}
where the functionals $C[N]$ and $C_z[N^z]$ are given by 
\begin{equation}\label{psConstr}
  \begin{eqtableau}{1}
    C[N] &:= \int dz\, N\, e^{-\gamma/2} \left[ 2\tau'' - 
      \gamma'\tau' - p_\gamma p_\tau + \tsfrac{p_\psi^2}{4\tau} + 
      \tau \psi'^2 \right] \\
    C_z[N^z] &:= \int dz\, N^z\, e^{-\gamma/2} \left[ p_\gamma \gamma' 
      + p_\tau \tau' + p_\psi \psi' - 2p_\gamma' \right]. \\
  \end{eqtableau}
\end{equation}
Note that, as one might have expected, the Hamiltonian of this system 
is written as a sum of constraints, and that the lapse and shift 
functions appear as Lagrange multipliers enforcing these constraints.

The phase space $\Gamma$ for our system is coordinatized by $(\gamma, 
p_\gamma, \tau, p_\tau, \psi, p_\psi)$.  To complete our description 
of $\Gamma$, we must specify the fall-off conditions on the canonical 
momenta.  These conditions will be motivated in essence by the 
requirement that the symplectic structure
\begin{equation}\label{bigSymp}
  \Omega[\delta_1, \delta_2] := \frac{1}{16\pi G} \int dz\, \left( 
    \delta_1p_\gamma\, \delta_2\gamma + \delta_1p_\tau\, \delta_2\tau 
    + \delta_1p_\psi\, \delta_2\psi - [1 \leftrightarrow 2] \right),
\end{equation}
as well as the Hamiltonian vector fields of the constraint functionals 
be well defined.  Furthermore, motivated by the spacetime treatment of 
this system in \cite{schmidt}, we hope eventually to use $\tau$ as the 
time parameter for the system.  We therefore expect that its time 
derivative should not vanish anywhere, even in the asymptotic regions.  
It follows from this, together with the definition of the momenta in 
terms of the time derivatives of the fields, that $p_\gamma$ cannot 
vanish asymptotically.  As stated in the previous subsection, however, 
we want to allow $\gamma$ to approach \textit{arbitrary} (non-zero) 
limits in the asymptotic regions.  The only way to ensure the 
convergence of the integral in eq.~\ref{bigSymp} is to require that 
$p_\gamma$ approach \textit{fixed} values in the asymptotic regions.  
In this way, although $p_\gamma$ will be finite asymptotically, we can 
guarantee that its \textit{variation} approaches zero rapidly enough 
to make the integral converge.  We therefore take the fall-off 
conditions for both the fields and their momenta to be
\begin{equation}\label{psFall}
  \begin{eqtableau}{2}
    \gamma   &\to \gamma_\pm(t) + \order(z^{-1}) &\qquad
    p_\gamma &\to -1 + \order(z^{-2})            \\
    \tau     &\to \tau_\pm(t) + \order(z^{-1})   &\qquad
    p_\tau   &\to \order(z^{-2})                 \\
    \psi     &\to \order(z^{-1})                 &\qquad
    p_\psi   &\to \order(z^{-1})                 \\
    N        &\to N_\pm(t) + \order(z^{-1})      &\quad
    N^z      &\to \order(z^{-1}).                \\
  \end{eqtableau}
\end{equation}
The asymptotic value $p_\gamma \to -1$ is chosen here primarily to 
simplify the discussion of gauge fixing in the next subsection.

Let us return now to the discussion of the fall-off of the shift function 
begun in the previous subsection.  In phase space terms, $C_z[N^z]$ 
represents a class of functions on phase space which we require to be 
differentiable, and the fall-off condition on $N^z$ amounts to a 
restriction on this class of functions.  The phase space functions 
corresponding to shifts which do not vanish asymptotically are not 
differentiable at \textit{any} point of the phase space described above.  
To have them be differentiable, we would have to work on a different phase 
space wherein $\gamma$ would also approach fixed values in the asymptotic 
regions.  Since the class of $N^z$ which vanish asymptotically already 
separates the points of a spatial slice (and can therefore enforce the 
local constraint \textit{density}), there is no need to use the larger 
class of $N^z$ which are free to take arbitrary asymptotic values.  Thus, 
our choice of fall-off conditions on the shift function is not only 
sufficient to enforce the constraint, but also preferable since the 
constraint functional is then defined on a phase space which can describe 
many more spacetimes.

We now have a complete characterization of the phase space $\Gamma$ and 
the constraint functionals $C[N]$ and $C_z[N^z]$.  A straightforward 
calculation shows that these constraints are first class.  We have seen in 
eq.~\ref{psAct} that the Hamiltonian for our system can be written as a 
sum of constraints and therefore vanishes identically on the constraint 
surface.  Note that this is \textit{not} the case in the canonical 
treatment of the Einstein-Rosen model in \cite{pier1}.  In that model, the 
Hamiltonian was equal to the surface term in the gravitational action, 
eq.~\ref{stdAct}.  In the present model, this surface term vanishes due to 
the boundary conditions \ref{psFall} we have chosen for the fields.  As a 
result, there is no canonical separation of gauge transformations and true 
dynamical evolution in our system as there was in the Einstein-Rosen case.  
To accomplish this separation and isolate the true degrees of freedom of 
the theory, we are forced to break the space-time covariance in a more or 
less ad hoc way.  We will do so in the following subsection by using 
intuition garnered from the spacetime picture to ``deparameterize'' the 
system \cite{depar}.

\subsection{Deparameterization}

There are a number of ways to approach the phase space reduction of a 
generally covariant system (e.g., frozen time formalism \cite[and 
references therein]{frozen}, presymplectic mechanics \cite{carlo}, etc.).  
The simplest approach in the present case seems to be that of 
deparameterization \cite{depar}.  The idea of the deparameterization 
procedure is that a generally covariant system must contain its own time 
parameter.  There is a (highly non-unique!)  procedure to isolate this 
time parameter and reduce the phase space to a proper symplectic manifold.

We begin with a presymplectic space $(\Gamma, \Omega)$ on which there are 
a number of first class constraints.  First, we seek to gauge-fix all but 
one of the first class constraints.  The gauge-fixed constraint surface 
$\hat\Gamma$ will then be odd dimensional, and the pullback of the 
presymplectic form to this surface will have precisely one degenerate 
direction given by the Hamiltonian vector field $X_C$ of the remaining 
constraint.  Second, we need to find a phase space function $T$ which 
satisfies $X_C(T) = 1$.  The level surfaces of $T$ will be everywhere 
transverse to $X_C$, so the pullback $\OmBar$ of $\Omega$ to any one level 
surface $\GamBar$ will be non-degenerate.  It follows that $(\GamBar, 
\OmBar)$ is a true symplectic space.  This is the phase space of the 
deparameterized theory.  Note that, in the end, some of the observables we 
will want to consider may be time-dependent.  Therefore, the observables 
of the reduced theory should be given by equivalence classes of functions 
which agree on $\hat\Gamma$, rather than just on $\GamBar$.  This 
asymmetry in the treatment of the constraints when defining observables 
seems to be where the most injustice is done to the original covariance of 
the system.  On the other hand, the deparameterization procedure yields a 
concrete model in which it is possible to calculate physically interesting 
quantities.  Obviously, this is a very strong argument in its favor.

Let us apply these ideas to the phase space described in the previous 
subsection.  In this particular situation, we already know from the 
spacetime formulation of the theory that we would like to identify $\tau$ 
with the parameter time $t$ of the system.  Therefore, we choose the gauge 
fixing conditions to be
\begin{equation}\label{gFix}
	\tau' = 0 \quad\mbox{and}\quad p_\gamma + 1 = 0.
\end{equation}
The first of these conditions guarantees that the prospective time 
parameter is constant on a spatial slice of the dynamical foliation of 
spacetime.  The second essentially does the same for $\dot\tau$.  Note 
that the second condition is actually equivalent to $p_\gamma' = 0$ since 
we required $p_\gamma$ to approach $-1$ at its asymptotic limits.  
However, we made no such requirement of $\tau$.  Thus, although $\tau$ is 
constant on the spatial slice, its value may still vary in time.  In other 
words, the infinite number of degrees of freedom represented by $\tau(z)$ 
have been reduced to just one, whereas those represented by $p_\gamma$ 
have been eliminated altogether.

To check that eqs.~\ref{gFix} are admissible gauge fixing conditions for 
use in the deparameterization program, we need to compute their Poisson 
brackets with the constraints:
\begin{equation}
	\begin{eqtableau}{2}
		\pbrack{\tau'}{C[N]} &\weq \left[ Ne^{-\gamma/2} \right]' &\qquad 
		\pbrack{\tau'}{C_z[N^z]} &\weq 0 \\
		\pbrack{p_\gamma + 1}{C[N]} &\weq 0 &\qquad \pbrack{p_\gamma + 
		1}{C_z[N^z]} &\weq -\left[ N^z e^{-\gamma/2} \right]'.  \\
	\end{eqtableau}
\end{equation}
Since the matrix so defined is invertible, the gauge-fixing conditions are 
indeed admissible except when
\begin{equation}
	N = N_0 e^{\gamma/2} \quad\mbox{or}\quad N^z = N_0^z e^{\gamma/2},
\end{equation}
where $N_0$ and $N_0^z$ are arbitrary constants.  Note, however, that the 
second possibility here is ruled out by the fall-off conditions imposed on 
$N^z$.  Thus, as required, there is exactly one first class constraint 
\textit{function} $C$ (up to scaling) which is not solved by these gauge 
fixing conditions.  Observe that setting $N_0 = 1$ will give $X_C(\tau) = 
1$.  This choice will therefore allow us to identify $\tau$ with the 
parameter time $t$ of the system as expected.

Let us now turn to the characterization of the phase space of the 
deparameterized theory.  To describe the surface $\GamBar$, we must 
first solve the set of second class constraints formed by the first 
class constraints, eqs.~\ref{psConstr}, together with their 
gauge-fixing conditions, eqs.~\ref{gFix}.  When we then identify 
$\tau$ with the parameter time $t$ of the theory, we 
find\footnote{Note that we have made an additional choice in the 
expression for $\gamma$.  The solution of the constraint density of 
$C[N]$ actually only implies that $\gamma' = p_\psi \psi'$.  To 
recover $\gamma$, we have to integrate this equality, leading to an 
undetermined constant of integration.  This constant can be used to 
fix the value of $\gamma$ at any one point of the spatial slice to be 
any value we like.  We have chosen to take $\gamma \to 0$ as $z \to 
-\infty$.  This choice can be thought of as the completion of the 
gauge fixing procedure outlined above.}
\begin{equation}\label{solConstr}
  \begin{eqtableau}{2}
    \gamma   &= \int_{-\infty}^z dz'\, p_\psi \psi' &\qquad
    p_\gamma &= -1                                  \\
    \tau     &= t                                   &\qquad
    p_\tau   &= -\tsfrac{p_\psi^2}{4t} - t \psi'^2. \\
  \end{eqtableau}
\end{equation}
These calculations show that $\GamBar$ is coordinatized by the pair 
$(\psi, p_\psi)$.  All the other dynamical variables of the original 
system are redundant and should now be treated as observables of the 
reduced system.  Thus, we see that all the true degrees of freedom of 
our theory now reside in the scalar field and its conjugate momentum.  
This agrees with the usual notion that in 2+1-dimensional gravity, all 
the local degrees of freedom should reside in the matter fields.

To finish the construction of the deparameterized phase space, we also 
have to specify the symplectic structure and the Hamiltonian function.  
Both of these can be found by restricting the action functional to 
$\GamBar$.  We find the reduced action to be
\begin{equation}\label{redAct}
  \SBar = \frac{1}{16\pi G} \int dt\, \left( \int dz\, \left[ 
    p_\psi \dot\psi \right] - \int dz\, \left[ \tsfrac{p_\psi^2}{4t} + 
    t \psi'^2 \right] \right).
\end{equation}
The first term of this action gives us a canonical 1-form on $\GamBar$ 
whose exterior derivative is the reduced symplectic structure 
$\OmBar$.  Clearly, we will just find that $\psi$ and $p_\psi$ are 
canonical coordinates on the reduced phase space.  The second term in 
the action gives the Hamiltonian for the deparameterized theory.  
Remarkably, these agree exactly with the symplectic structure and 
Hamiltonian of a scalar field theory on a \textit{fixed} background 
spacetime whose metric is given by
\begin{equation}\label{bgMet}
  \overcirc g_{ab} = -\grad_a t \, \grad_b t + \grad_a z \, \grad_b z 
    + t^2 \grad_a \sigma \, \grad_b \sigma.
\end{equation}
This metric is again defined on a 3-manifold with topology $\Re^2 
\times S^1$.  In fact, this background spacetime corresponds to the 
point $\psi = p_\psi = 0$ of the reduced phase space.

The identification of the reduced phase space for our system with that 
of a scalar field on a fixed background is very important for the 
quantization which we will describe in the following section.  This is 
due to the relative simplicity of the later description.  Although 
other descriptions of the reduced phase space are certainly possible, 
they do not offer as simple a quantization scheme as the scalar field.  
For this reason, it is useful to restrict our attention to the scalar 
field theory to study the quantization of our system.  We can then use 
the classical expressions derived above to find quantum observables 
describing geometric quantities of interest.

\section{Quantum Theory}

\subsection{Preliminaries}

In the previous section, we were primarily concerned with the geometrical 
significance of the scalar field in our theory.  We therefore focussed on 
the dimensionless quantity $\psi$.  The approach we will take to 
quantization, however, is based on the fully reduced phase space described 
above --- that of a free scalar field.  Therefore, in this section, it is 
convenient to rescale the field so that it acquires the proper dimension 
for a physical scalar field in 2+1 dimensions.  Accordingly, we will now 
switch our attention to the quantity $\phi := \psi/\sqrt{8\pi G}$.  The 
natural choice for the momentum conjugate to $\phi$ is $p_\phi := 
p_\psi/\sqrt{32\pi G}$.  Using these variables, we can reexpress the 
reduced action of eq.~\ref{redAct} as
\begin{equation}\label{scalAct}
  S = \int dt\, \left( \int dz\, \left[ p_\phi \, \dot\phi \right] - 
    \int dz\, \tsfrac{1}{2} \left[ t^{-1} p_\phi^2 + t \phi'^2 \right] 
    \right).
\end{equation}
This is just the usual action for a free scalar field on the background 
spacetime given by eq.~\ref{bgMet}.  Thus, the phase space $\Gamma$ has 
the structure of a real vector space and is coordinatized by the canonical 
pair $(\phi, p_\phi)$.  Note that we have changed notation slightly from 
the previous section: we have dropped the bars over both the (reduced) 
action and the (reduced) phase space.  This is done to emphasize that the 
scalar filed theory is now regarded as fundamental.  The other geometrical 
quantities discussed above will now be treated as secondary, derived 
observables.

The quantization of scalar fields on fixed background spacetimes is by now 
very well understood.  In the present discussion, we will follow the 
standard constructions laid out, for example, in \cite{waldQFT}.  This 
procedure yields the Hilbert space of the quantum field theory by building 
a Fock space over a single-particle Hilbert space derived from the space 
$\mfs S$ of solutions of the classical equations of motion.  We should 
therefore begin with the description of this space.

The equations of motion for the field theory follow easily from the 
action \ref{scalAct}.  They are 
\begin{equation}\label{scalEoM}
  \left\{  \begin{eqtableau}{1}
    \dot\phi &= t^{-1} p_\phi \\ \dot p_\phi &= t\phi''
  \end{eqtableau}  \right\}  \quad\implies\quad 
  \ddot\Phi + t^{-1}\dot\Phi - \Phi'' = 0.
\end{equation}
Here, and below, the quantity $\Phi = \Phi(t, z)$ denotes a solution 
to the equations of motion, while $\phi = \phi(z)$ denotes the value 
of the field on a given spatial slice.  Note that, as expected for a 
free field theory, the equations of motion are linear.  It follows 
that the space $\mfs S$ of their real solutions will be a real vector 
space.  We may therefore expand the most general real-valued solution 
in terms of certain fundamental solutions as
\begin{eqset}\label{scalSol}
  \Phi(t, z) &= \int_{-\infty}^\infty \frac{dk}{\sqrt{2\pi}} \, \left[ 
    A(k) f_k^{(2)}(t, z) + \overline{A(k)} f_k^{(1)}(t, z) \right].  \\
\interject{with the fundamental solutions given by}
  f_k^{1 \choose 2}(t, z) &:= \frac{\sqrt\pi}{2} \, 
    H_0^{1 \choose 2}(\abs kt) \, e^{\mp ikz}, \\
\end{eqset}%
where $H_0^{1 \choose 2}(\cdot)$ denotes the zeroth-order Hankel 
function of type 1 or 2.  The wave profile $A(k)$ we have introduced 
here will generically be \textit{complex}, and can therefore be viewed 
as a complex coordinate on the real vector space $\mfs S$.  It will 
also have to satisfy come requirements related to the fall-off 
conditions on $\phi$ and $p_\phi$.  By allowing some modifications to 
the fall-off conditions given previously in eq.~\ref{psFall}, we can 
make these requirements precise.  In particular, if we require $\phi$ 
and $p_\phi$ to be Schwartz functions (i.e., they fall off at infinity 
faster than any polynomial), we will find the corresponding wave 
profiles are also Schwartz.  This change in fall-off conditions will 
not change any the results to follow owing to the completion of $\mfs 
S$ to a Hilbert space which will take place during the quantization.  
The point of all this is that we can concretely identify the solution 
space $\mfs S$ with the space of Schwartz functions of a single 
variable.

There is a natural map from the space $\mfs S$ to the phase space 
$\Gamma$ of our system.  One defines this map by identifying a 
solution of the equations of motion with its initial data at some 
fixed initial time $t_0$.  It is not difficult to show that this map 
is linear, invertible and (bi-)continuous when the fall-off conditions 
described above are applied.  We can therefore use it to induce a 
symplectic structure on $\mfs S$ using the given one on $\Gamma$.  
This symplectic structure can be written in terms of the wave profiles 
$A(k)$ as
\begin{equation}\label{wpSymp}
  \Omega(\Phi_1, \Phi_2) = i \, \int_{-\infty}^\infty dk \, \left[ 
    \overline{A_1(k)} A_2(k) - A_1(k) \overline{A_2(k)} \right].
\end{equation}
We have chosen to write the symplectic structure as an antisymmetric 
bilinear form on $\mfs S$ rather than as a differential 2-form.  There 
is no problem with this since $\mfs S$ is linear and can therefore be 
identified with its tangent space.  Moreover, the quantization 
procedure requires the symplectic structure be given as a bilinear 
form, rather than as a differential form, which explains why the 
procedure would not work for non-free field theories.

\subsection{Quantization}

The key step in the quantization of a free field theory is to pick a 
K\"ahler structure on the space $\mfs S$ of solutions to its classical 
equations of motion.  That is, we need to choose a complex structure 
$J: \mfs S \to \mfs S$ such that
\begin{equation}
  \mu(\Phi_1, \Phi_2) := -\tsfrac 12 \, \Omega(\Phi_1, J \circ \Phi_2)
\end{equation}
is a positive-definite real inner product on $\mfs S$.  There are 
always a number of possible ways to do this, and the various ways do 
not necessarily lead to equivalent quantum theories.  In our case, 
however, we already have a complex coordinate $A(k)$ on $\mfs S$, so 
there is a natural candidate for such a structure given by $J: A(k) 
\mapsto iA(k)$.  This is indeed an acceptable choice since it yields 
the inner product
\begin{equation}
  \mu(\Phi_1, \Phi_2) = \tsfrac 12 \int_{-\infty}^\infty dk\, \left[ 
    \overline{A_1(k)} A_2(k) + A_1(k) \overline{A_2(k)} \right], 
\end{equation}
which is manifestly real and positive-definite.

With the choice of a complex structure, it becomes possible to view 
$\mfs S$ as a \textit{complex} vector space.  Moreover, we can define 
a complex inner product on $\mfs S$ by 
\begin{equation}\label{iprod}
  \begin{eqtableau}{1}
    \iprod{\Phi_1}{\Phi_2}_1 &:= \tsfrac 1\hbar \, \mu(\Phi_1, \Phi_2) 
        - \tsfrac i{2\hbar} \, \Omega(\Phi_1, \Phi_2) \\
      &= \hbar^{-1} \int_{-\infty}^\infty dk\, \overline{A_1(k)} A_2(k), 
  \end{eqtableau}
\end{equation}
where the factors of $\hbar$ are included at this point to render the 
inner product of two vectors dimensionless.  This inner product will 
be Hermitian in the complex structure defined by $J$, and is again 
manifestly positive-definite.  Thus, we have endowed $\mfs S$ with the 
structure of a complex pre-Hilbert space.  The Cauchy completion of 
this space in the inner product norm gives a complex Hilbert space 
$\mfs H_1$ which, due to the simple form of the inner product 
$\iprod\cdot\cdot_1$, can immediately be identified with $L^2(\Re)$.  
This is the single-particle state space for the quantum theory.  The 
Hilbert space of the field theory is then given by the symmetric Fock 
space $\mfs H := \mfs F_S \left( \mfs H_1 \right)$ built on $\mfs 
H_1$.

To complete the quantization of the theory, we have to introduce a set 
of operators on $\mfs H$ corresponding to a ``complete'' set of 
classical observables and specify a Hamiltonian operator.  It is 
simplest to do the first of these in the form of the quantum field 
itself 
\begin{equation}
  \hat\Phi(t, z) := \sqrt\hbar \int_{-\infty}^\infty 
    \frac{dk}{\sqrt{2\pi}} \, \left[ f_k^{(2)}(t, z) \, \hat a_k + 
    f_k^{(1)}(t, z) \, \hat a_k^\dagger \right], 
\end{equation}
which we have expressed using the creation and annihilation operators 
provided by the Fock structure of the Hilbert space.  As usual, this 
is actually an operator-valued distribution.  The true observables of 
the theory are the smeared field operators $\hat\phi(t; g]$ and $\hat 
p_\phi(t; g]$ gotten by integrating this distribution (and the one 
corresponding to its canonical momentum) against a suitably 
well-behaved function $g(z)$.  One can check that the smeared field 
operators do actually obey the proper canonical commutation relations.

The classical Hamiltonian of the system is given by the second term in 
the action \ref{scalAct}.  It can be promoted to a quantum observable 
on $\mfs H$ which will take the standard form
\begin{equation}
  \hat H = \int_{-\infty}^\infty dk\, \hbar\abs k a_k^\dagger a_k.
\end{equation}
We have chosen the conventional normal-ordering of the right side of 
this expression to subtract the (infinite) ground state energy and 
make $\hat H$ a well-defined operator on the Hilbert space $\mfs H$.  
It does not require any further regularization.

We conclude this discussion with a remark.  As we noted above, there 
is no canonical choice of complex structure on $\mfs S$ and different 
choices can lead to inequivalent quantum theories.  In the case of 
Minkowskian quantum field theories, the complex structure is fixed by 
the requirement of Poincar\'e invariance.  In the present case, 
however, we have only a rotational invariance which unfortunately is 
insufficient to determine the complex structure uniquely.  As a 
result, one should note that the quantization we have performed above 
is \textit{not} unique and there are other realizations of the quantum 
system.  Nevertheless, the complex structure we have chosen does have 
some nice properties which justify its use, even though they do not 
single it out.  From the geometric point of view, the only interesting 
classical observable of the system is the metric function $\gamma$.  
This function is quadratic on the phase space of the system but 
unfortunately is not bounded from below.  The complex structure we 
have used here does commute with the infinitesimal canonical 
transformation generated by $\gamma$ on the classical phase space.  
Had $\gamma$ been bounded from below, this commutativity would have 
been sufficient to choose the complex structure uniquely 
\cite{ashsen}.  As it stands, this fact only shows that our choice is 
one of a class of complex structures which are well adapted to the 
observables of interest for the system.  From a practical point of 
view, however, this choice of complex structure has the property that 
it \textit{does} allow us to complete the quantization of the model 
exactly.

\subsection{Quantum Geometry}

Since in the previous two subsections we have constructed an exact quantum 
theory of a gravitational system, we can now ask whether there are new 
physical insights to be drawn from the model.  In this subsection, we will 
see that there are indeed some lessons we can learn.  In \cite{ashLQE}, it 
was shown that surprisingly large quantum dispersions in the Coulombic 
modes of the spacetime metric would result from the presence of 
high-frequency excitations in the quantum Einstein-Rosen model of 
\cite{pier1}.  The Schmidt model is similar to the Einstein-Rosen model in 
certain respects, but has a decidedly different overall structure.  It is 
therefore natural to ask whether these large quantum effects persist in 
this model.  As we will see below, they do.  This will constitute a first 
check on the robustness of the dispersion results.

The content of the spacetime geometry in the Schmidt model is encoded in 
the 2+1-dimensional metric $g_{ab}$.  In the classical theory, after the 
gauge-fixing conditions have been applied, the metric of eqs.~\ref{2+1met} 
and \ref{1+1met} becomes
\begin{equation}
  g_{ab} = e^{\gamma(t, z)} (-\grad_a t \, \grad_b t + 
    \grad_a z \, \grad_b z) + t^2 \grad_a \sigma \, \grad_b \sigma,
\end{equation}
where $\gamma(t, z)$ is defined in terms of the scalar field by 
eq.~\ref{solConstr}.  The only nontrivial component of this metric is 
clearly $g_{zz} = -g_{tt} = e^{\gamma(t, z)}$, and we will accordingly 
focus our attention here on the quantum analog $e^{\hat\gamma(z, t)}$ of 
this observable.

Let us begin by describing the operator $\hat\gamma(t, z)$ itself.  We 
can use the classical expression of eq.~\ref{solConstr} to express 
this in terms of the creation and annihilation operators as 
\begin{eqset}\label{finGamma}
  \hat\gamma(t, z) &= 16\pi G\hbar \int dk\, d\ell\, \Bigl[
    \overline{\Delta_+(t, z; k, \ell)}\, 
      \hat a_k^\dagger \hat a_\ell^\dagger + 
    \Delta_+(t, z; k, \ell)\, \hat a_k \hat a_\ell + {} \nonumber\\
    &\hphantom{{}= 16\pi G \int dk\, d\ell\, \Bigl[} \qquad 
    \Delta_-(t, z; k, \ell)\, \hat a_k^\dagger \hat a_\ell + 
    \overline{\Delta_-(t, z; k, \ell)}\, 
      \hat a_\ell^\dagger \hat a_k \Bigr] \\
\interject{where the Green's functions $\Delta_\pm(t, z; k, \ell)$ are 
  given by}
  \Delta_\pm(t, z; k, \ell) &= \frac{i\abs k\ell t}{8} \, 
    H_1^{2 \choose 1}(\abs kt) \, H_0^{(2)}(\abs\ell t) 
    \int_{-\infty}^z dz'\, e^{i(\ell \pm k)z'}. \label{green}\\
\end{eqset}%
Note that, as with the Hamiltonian above, we have chosen to 
normal-order the right hand side of eq.~\ref{finGamma}.  This is done 
both to regularize the operator and to preserve the classical 
relationship between $\gamma$ and the field momentum.  Also note that 
the Green's functions $\Delta_\pm(t, z; k, \ell)$ which we use here 
are actually distributional due to the last term in eq.~\ref{green} 
(which is the Fourier transform of a step function).  Therefore, 
eq.~\ref{finGamma} does not define an operator on the Hilbert space 
$\mfs H$, but rather another operator-valued distribution.  To make 
sense of this as a proper observable, one must define an appropriate 
class of smearing functions to mitigate the singular nature of the 
integrals involved.  Although this is certainly possible in principle, 
there are significant technical problems with the procedure; we will 
concentrate instead on the structure of the \textit{asymptotic} metric 
operator.  This is both technically simpler and physically more 
interesting.

All of the spacetimes we consider in this paper have two distinct 
asymptotic regions.  In each of these regions, the expression for the 
operator $\hat\gamma$ of eq.~\ref{finGamma} simplifies considerably.  
In the limit as $z \to -\infty$, we find that the asymptotic form of 
$\hat\gamma$ is identically zero.  Therefore, the metric in this 
region will simply agree with the background metric introduced in 
eq.~\ref{bgMet}.  In particular, it is a ``c-number;'' it has no 
dispersion.  This is really a result only of the choices we made in 
the gauge fixing procedure.  Recall that, classically, we were able to 
fix the value of $\gamma$ to be any given quantity at any particular 
point $z$, and that we used this freedom to fix $\gamma \to 0$ as $z 
\to -\infty$.  The quantum mechanical theory, then, is based on a 
space where other asymptotic values of $\gamma$ simply do not occur.  
In light of these facts, the above result about the metric operator 
$\hat\gamma_-$ is not surprising.  It is also not very interesting.

In the opposite asymptotic limit, as $z \to +\infty$, the 
distributional integrals in eq.~\ref{green} converge, in an 
appropriate sense, to delta functions.  As a consequence, the 
asymptotic form of $\hat\gamma$ becomes
\begin{equation}\label{asympGam}
  \hat\gamma_+  := \lim_{z \to \infty} \hat\gamma(t, z)
    = -16\pi G \int dk\, \hbar k \, a_k^\dagger a_k.
\end{equation}
Unlike the generic expression for $\hat\gamma(t, z)$ given in 
eq.~\ref{finGamma}, this expression \textit{does} define a proper 
self-adjoint operator on $\mfs H$.  Thus, we may define and study the 
geometrical operator $\hat g_{zz}^+ = e^{\hat\gamma_+}$.

To illustrate the quantum gravitational effects in the system, we wish 
to concentrate on a class of semi-classical states which approximate 
classical spacetimes.  Luckily, there already exists a well-known 
class of semi-classical states which approximate classical states of 
the scalar field, namely the coherent states
\begin{equation}\label{coState}
  \ket A := e^{-\norm A_1^2/2} \, e^{\hbar^{-1/2} \int dk\, A(k)\, 
    \hat a_k^\dagger} \ket 0.
\end{equation}
The norm $\norm \cdot_1^2$ which appears here is that associated with 
the one-particle Hilbert space inner product of eq.~\ref{iprod}, and 
$\ket 0$ denotes the (unique) ground state of the theory.  As usual, 
each of these states is peaked at the classical configuration of the 
scalar field given by the wave profile $A(k)$ and minimizes the 
uncertainty in the value of the field.  With the normalization factor 
given in eq.~\ref{coState}, this set of coherent states also reflects 
both the classical and quantum structure of the system:
\begin{equation}
  \braket{A_1}{A_2} = e^{-\norm{A_1 - A_2}_1^2/2} \, 
    e^{-i\Omega(A_1, A_2)/2\hbar}. 
\end{equation}
The classical system is reflected in the phase of the inner product 
which involves the symplectic structure of eq.~\ref{wpSymp}.  On the 
other hand, the amplitude of the inner product is given in terms of 
the single-particle Hilbert space norm.  Although it has been shown 
\cite{gamPull} that there are other classes of semi-classical states 
which minimize the \textit{combined} uncertainty in the field and the 
metric, we will continue to use these here because of the close 
integration with the reduced classical system.

The results on quantum geometry we will obtain shortly all follow from 
a simple proposition.  Suppose we have an observable $\hat\mathcal O = 
\int dk\, f(k)\, \hat a_k^\dagger \hat a_k$ which can be written as a 
sum of contributions from each particle present in the quantum state.  
The action of the exponential of this operator on a coherent state of 
the particle system is then given by 
\begin{equation}\label{expOp}
  e^{\hat\mathcal O} \ket A = \exp \left( \tsfrac 1{2\hbar} \int dk\, 
    \left[ e^{2f(k)} - 1 \right] \, \abs{A(k)}^2 \right) \ket{e^f A}, 
\end{equation}
where $\ket{e^f A}$ denotes the coherent state associated with the 
wave profile $A'(k) = e^{f(k)} A(k)$.  Note that since, as we have 
done above, the wave profiles $A(k)$ are typically taken to be 
Schwartz functions, $\ket{e^f A}$ is not necessarily a well-defined 
coherent state.  If $f$ diverges faster than logarithmically at 
infinity, the function $e^f A$ will not necessarily be Schwartz, and 
the operator $e^{\hat\mathcal O}$ may not be defined.  This is not too 
surprising since the operator in question is certainly unbounded and 
we expect therefore that we should have to restrict the domain of 
definition of the operator.  Fortunately, there is a natural domain 
which we can always choose: the coherent states corresponding to wave 
profiles of compact support.  It is not difficult to show that these 
can approximate the Schwartz function coherent states to arbitrarily 
close precision.  Since the later are already (over-)complete, this 
shows the exponential operator can be densely defined.  Thus, although 
the expression may initially appear suspicious, we should not be too 
concerned with the functional analytic subtleties of eq.~\ref{expOp}.

\newcommand{\metExpect}{\expect{\hat g_{zz}^+}}
\newcommand{\metUncert}
  {\frac {\left( \Delta g_{zz}^+ \right)^2} {\expect{g_{zz}^+}^2}}

Since, according to eq.~\ref{asympGam}, $\hat\gamma_+$ is of the 
required form, we can apply the above result to the study of the 
metric operator $\hat g_{zz}^+$.  The expectation value of the metric 
operator and the relative uncertainty in its measurement may be 
expressed in closed form as
\begin{eqset}\label{expect}
  \metExpect &= \exp \left( \hbar^{-1}\, \int dk\, \left[ 
    e^{-16\pi G\hbar k} - 1 \right] \, \abs{A(k)}^2 \right) \\ \label{uncert}
  \metUncert &= \exp \left( \hbar^{-1}\, \int dk\, \left[ 
    e^{-16\pi G\hbar k} - 1 \right]^2 \, \abs{A(k)}^2 \right) - 1. \\
\end{eqset}
Meanwhile, the classical expression for the asymptotic metric is 
\begin{equation}\label{class}
	\left( g_{zz}^+ \right)_\mathrm{classical} = \exp \left( -16\pi G 
  	\int dk\, k\, \abs{A(k)}^2 \right).
\end{equation}
These expressions clearly show that there are non-trivial quantum 
effects present in our system, even when the scalar field is sharply 
peaked at a classical configuration.  Since we now have a canonical 
frequency scale given by the Planck value $1/G\hbar$, we can gain some 
qualitative understanding of these effects by examining some of the 
limiting behavior of these expressions.  In each case, we will 
consider a wave profile $A(k)$ which is sharply peaked at a certain 
characteristic wave number $k_0$.  There are three distinct regimes we 
will discuss.  In the discussion, it is useful to define $N := \norm 
A_1^2$, which may be interpreted as the expected total number of 
particles in the coherent state associated with the wave profile 
$A(k)$.

\begin{enumerate}

\item\label{lf} \textit{Low frequency} ($\abs{G\hbar k_0} \ll 1$): In 
this case, we can simply expand the various expressions in powers of 
$G\hbar k_0$ to find the expectation value of the metric operator and 
its relative uncertainty.  The results are:
\begin{equation}
  \begin{eqtableau}{1}
  	\metExpect &\approx \left( g_{zz}^+ \right)_\mathrm{classical}\, 
	    \left[ 1 + \tsfrac N2 (16\pi G\hbar k_0)^2 + \cdots \right] \\
    \metUncert &\approx N (16\pi G\hbar k_0)^2 + \cdots. \\
  \end{eqtableau}
\end{equation}
This shows that in the low frequency limit, both the deviation of the 
quantum metric from the classical and its relative uncertainty are 
second order in the expansion parameter\footnote{Note that we should 
also have $N(G\hbar k_0)^2 \ll 1$ to get these results.  As a result, 
the number of particles expected in the quantum state cannot be too 
large.  This limitation is somewhat surprising since the coherent 
states of the scalar field approximate their classical counterparts 
only when the expected particle number \textit{is} large: $N \gg 1$.  
However, since the characteristic frequency of the particles, $G\hbar 
k_0$, is already very small, there is presumably some intermediate 
range wherein both approximations are reasonably accurate.}.  Thus, in 
this regime, the quantum system closely mimics the classical one.

\item\label{hff} \textit{High frequency, forward direction} ($G\hbar 
k_0 \gg 1$): This situation occurs when the quantum state describes a 
number of high-frequency particles moving \textit{toward} $z = 
+\infty$.  In this limit, the expectation value of the metric and its 
relative uncertainty become
\begin{equation}
	\metExpect \approx e^{-N} \quad\mbox{and}\quad 
	\metUncert \approx e^N - 1
\end{equation}
Meanwhile, the classical metric in this limit becomes $e^{-16N\pi 
G\hbar k_0}$, which is small, even when compared with the quantum 
expectation value.  Furthermore, when there are more than a couple 
particles present, the uncertainty in the metric can be quite large 
compared to its expectation value.

\item\label{hfb} \textit{High frequency, backward direction} ($-G\hbar 
k_0 \gg 1$): This situation occurs when the quantum state describes a 
number of particles moving \textit{away from} $z = +\infty$.  In this 
limit, we can again find approximate values for the metric and its 
relative uncertainty
\begin{equation}
	\metExpect \approx \exp \left( Ne^{-16\pi G\hbar k_0} \right) 
	\quad\mbox{and}\quad 
	\metUncert \approx \exp \left( Ne^{-32\pi G\hbar k_0} \right).
\end{equation}
The classical metric in this case is again approximately $e^{-16N\pi 
G\hbar k_0}$.  This, too, is small compared to the quantum expectation 
value.  In this case, even when there is only \textit{one} particle 
expected in the quantum state, the uncertainty in the metric can be 
huge compared to its expected value.

\end{enumerate}

These results seem to agree, on the whole, with the results found in 
\cite{ashLQE} regarding the cylindrical wave case.  There \textit{do} 
exist semi-classical states within our system, but they belong to a 
very restrictive class.  Specifically, states with too few or too 
many particles, as well as \textit{any} state which contains particles 
of Planckian or trans-Planckian frequencies do not occur in the 
classical limit of the quantum theory.  All of these states are 
narrowly peaked around a particular classical field configuration.  
However, in the case of the excluded states, not only is the 
uncertainty in the metric large, but the expected value of the metric 
is wildly different from the classical approximation.  It has been 
shown in \cite{gamPull} that the uncertainty in the metric measurement 
is partially a consequence of the particular family of semi-classical 
states we used in our analysis; a different choice could decrease this 
uncertainty at the expense of introducing a larger uncertainty in the 
value of the field.  However, even if some other set of states were 
chosen, the corresponding classical field configurations would remain 
poor approximations to the solutions of the full, interacting quantum 
theory.

\section{Discussion}

It has been known for some time that general relativity, under the 
assumption of two hypersurface-orthogonal Killing fields, can be described 
in terms of a free scalar field theory.  In light of this fact, the 
equations governing the Schmidt model are not unexpected.  However, in the 
phase space formulation and quantization, there \textit{are} some unusual 
aspects of the construction which merit further discussion.

At the classical level, we considered a class of spacetimes with the 
somewhat unusual property that they are neither spatially compact nor 
asymptotically flat.  These properties, one of which is usually 
assumed for one reason or another, often play a decisive role in 
formulating the canonical theory for the system under consideration.  
Nevertheless, we saw that it is possible to define a proper phase 
space for the Schmidt model by requiring only the convergence of the 
symplectic structure \ref{bigSymp} and the differentiability of the 
constraint functions \ref{psConstr}.  However, the resulting classical 
system does not have any natural time structure and its Hamiltonian 
vanishes weakly.  To isolate its true, independent degrees of freedom, 
we therefore found it necessary to ``deparameterize'' the theory.  
Fortunately, since we had previous knowledge of the spacetime 
treatment of the model, we had a natural candidate for the time 
parameter of the reduced system.  Furthermore, after the reduction, 
the system took the expected form of a free scalar field on a fixed 
background.  This was the key fact which enabled us to quantize the 
theory in a relatively straightforward way.

In the quantum theory, we saw that there is a natural class of 
semi-classical states of the quantum system which are built from its 
classical solutions.  These states can be identified with the usual 
coherent states of the scalar field theory which describes the reduced 
model.  As was pointed out in \cite{gamPull}, there are ways of 
realizing other such sets of semi-classical states within the quantum 
theory.  However, the set which we have chosen is particularly well 
adapted to a discussion of the modifications to the fixed-background 
scalar field theory which result from coupling it to a dynamical 
gravitational field.  This is because these semi-classical states 
remain sharply peaked at a classical scalar field configuration even 
in the presence of the gravitational field.  The other possible sets 
of coherent states do minimize a \textit{combination} of the metric 
and scalar field uncertainties at the expense of introducing a larger 
dispersion in the value of the scalar field by itself.

Using the exact quantum model we had constructed, we were able to find 
some unexpectedly large dispersions in the (asymptotic) metric caused 
by the presence of high-frequency scalar field excitations.  
Surprisingly, the dispersions are independent of where the particle 
is located.  Even a single high-frequency particle located at any 
point of space can cause very large dispersions in the 
\textit{asymptotic} metric.  This result can be interpreted at a 
couple of levels.

Firstly, one can consider the implications for quantum field theory in 
general.  It is often hypothesized that the introduction of gravity will 
introduce a natural cut-off in quantum field theory that will obviate the 
need for renormalization.  In our model, no such cut-off has emerged.  
However, the dispersions in the metric expectation suggest that a 
classical spacetime geometry simply fails to be a good approximation to 
the quantum state when high-frequency particles are present\footnote{There 
is another possible interpretation of this result: that the dispersions 
signal the breakdown of the rotational symmetry of the configuration 
rather than of the spacetime picture itself.  That is, that symmetric 
high-frequency excitations of the fields are unstable and rapidly become 
asymmetric when quantum effects are taken into account.  By itself, 
however, this would be an unforeseen, genuinely quantum mechanical effect.  
There are efforts under way \cite{asym} to examine this question by 
studying quantum gravity effects in configurations which are 
asymmetrically perturbed from the symmetric states described here.}.  The 
coupling of the matter field to the gravitational field seems to cause the 
classical spacetime structure to break down in this regime.  Conversely, 
the absence of metric dispersions when no high-frequency particles are 
present suggests that flat spacetime quantum field theory is a reasonable 
approximation to the true situation in that regime.  These facts lend some 
concrete credence to the notion that quantum field theory is a low-energy 
limit of a larger theory which includes gravity.  As such, infinities can 
and do arise when this limiting theory is pushed beyond its domain of 
validity.

The Schmidt model also provides an infinite number of examples of 
exact solutions to semi-classical gravity.  Semi-classical gravity is 
the theory of quantum fields propagating on a \textit{dynamical} 
classical spacetime.  A state of this theory must therefore specify a 
Lorentzian manifold $(M, g_{ab})$, a quantum field $\hat\phi$ on that 
manifold, and a state $\ket\psi$ of the quantum field which satisfy 
the dynamical equation
\begin{equation}\label{semiClass}
  G_{ab} = 8\pi\expect{\hat T_{ab}}_\psi.
\end{equation}
In our model, the expectation value of the stress-energy operator in any 
coherent state is exactly equal to its value in the associated classical 
state.  Consequently, eq.~\ref{semiClass} will be satisfied for any 
coherent state by taking the metric to be that of the corresponding 
classical solution.  However, as was pointed out in \cite{ashLQE}, the 
states of semi-classical gravity which include high-frequency particles 
should not be taken seriously.  Again, these solutions do not approximate 
the full quantum gravitational theory at all closely.

Finally, we should discuss the limitations of this model.  At a technical 
level, we have described only one possible quantization of the Schmidt 
model; it is not unique.  There are two points at which one could make 
different choices to quantize the system.  First, as was discussed above, 
it is possible to pick a different complex structure to use in the 
quantization procedure.  Although different choices \textit{can} lead to 
inequivalent quantum theories, this ambiguity is fairly tame for our 
purposes since the large quantum gravity effects which we have found in 
this paper will persist in alternate quantizations.  Second, and more 
importantly, there are ambiguities in the gauge fixing procedure already 
at the classical level.  To make sense of our theory for quantization, we 
had to isolate its true degrees of freedom using a deparameterization 
procedure.  In this case, it is \textit{not} clear that alternate gauge 
fixings would reproduce the qualitative content of our results.  However, 
from a practical point of view, one should note that the quantization 
procedure given our choices \textit{can} be completed and that we 
\textit{can} recover concrete results from the theory.  In general, this 
would probably not be the case.

At a more fundamental level, although we have caught some glimpses of 
quantum geometry in this model, the spacetimes involved are only 
2+1-dimensional.  The physically interesting case, of course, is that of 
3+1-dimensional gravity, and there are fundamental differences between the 
two.  Most notably, the 2+1-dimensional spacetimes do not allow the 
possibility of black holes (without permitting a non-zero cosmological 
constant).  The present system models a different sector of the 
3+1-dimensional theory: the radiative modes of the gravitational field 
which are not sufficiently strong to cause gravitational collapse.  The 
simplest 3+1-dimensional analog of the systems discussed here and in 
\cite{pier1, pier2} describes the collapse of a spherically symmetric 
scalar field in a Schwarzschild spacetime.  It would be quite illuminating 
to understand the quantization of that model.  There are at least two 
reasons to expect this.  First, it incorporates the black hole sector of 
gravity which the models considered until now have ignored.  Second, it is 
a truly 3+1-dimensional system, so the radiative modes of the 
gravitational field are expected to die off asymptotically as the inverse 
of the radius rather than logarithmically.  Consequently, the details of 
the dispersion effects due to high-frequency excitations may differ 
slightly from those described here.  Nevertheless, since all of these 
models describe scalar fields propagating on spacetimes on which the 
gravitational field does not have its own local degrees of freedom, it is 
reasonable to hope that the calculations in the spherically symmetric case 
can be completed using intuition garnered from the lower-dimensional 
models.

\topic{Acknowledgments}

I would like to thank Laurent Friedel, Raymond Puzio, Lee Smolin and 
especially Abhay Ashtekar for helpful discussions.  This work was 
supported by the NSF grant PHY-95-14240 and by the Eberly 
research funds of the Pennsylvania State University.

\end{document}